\documentclass[%
twocolumn,
 fleqn,usenatbib,
 showkeys,
 floatfix,nolongbibliography,aps,
author-numerical%
]{revtex4-2}

\pdfoutput=1
\usepackage[T1]{fontenc}
\usepackage [latin1]{inputenc}
\DeclareRobustCommand{\VAN}[3]{#2}
\let\VANthebibliography\thebibliography
\def\thebibliography{\DeclareRobustCommand{\VAN}[3]{##3}\VANthebibliography}

\usepackage{newtxtext,newtxmath}
\usepackage{tikz,xcolor,hyperref}
\usepackage[normalem]{ulem}
\usepackage{graphicx,url,caption,subcaption}
\usepackage{dcolumn}
\usepackage{bm}

\usepackage{xcolor}

\definecolor{lime}{HTML}{A6CE39}
\DeclareRobustCommand{\orcidicon}{%
	\begin{tikzpicture}
	\draw[lime, fill=lime] (0,0) 
	circle [radius=0.16] 
	node[white] {{\fontfamily{qag}\selectfont \tiny ID}};
	\draw[white, fill=white] (-0.0625,0.095) 
	circle [radius=0.007];
	\end{tikzpicture}
	\hspace{-2mm}
}

\foreach \x in {A, ..., Z}{%
	\expandafter\xdef\csname orcid\x\endcsname{\noexpand\href{https://orcid.org/\csname orcidauthor\x\endcsname}{\noexpand\orcidicon}}
}

\begin{document}
\title{Peristaltic Flow in Compressible, Ideal Magnetohydrodynamics: A Mechanism For Solar Spicules}
\author{D. Tsiklauri\orcidA{}}
 \email{D.Tsiklauri@salford.ac.uk}
\affiliation{Joule Physics Laboratory,
School of Science, Engineering and Environment, 
University of Salford,
Manchester, M5 4WT, 
United Kingdom}
\date{\today}

\begin{abstract}
We present an analytical model for peristaltic transport within compressible, ideal magnetohydrodynamics (MHD). By employing a small-amplitude perturbation expansion, under a thin-tube long-wavelength approximation with a uniform axial background magnetic field, we study non-linear coupling between thermodynamic pressure variations and Maxwell's magnetic tension stresses. The resulting net time-averaged conserved mass flux $\langle \dot{M} \rangle$ is calculated. When applied to solar chromospheric spicules under equipartition constraints ($\beta \sim 1$), where sound speed matches the Alfv{\'e}n speed, we find $\langle \dot{M} \rangle = 8\epsilon^2/(M^2-1)^2$. Because the denominator remains strictly positive across all operational supersonic Mach numbers ($M \approx 2\text{--}10$), upward-propagating mechanical disturbances drive a highly directional, collimated upward flow which we interpret as a spicule.
Estimates show that for observationally realistic magnetosonic waves with amplitudes of $\approx 10\%$, the peristaltic mechanism generates a localized mass flux up to $\approx 100$ times that of the solar wind. We show that because high-Mach waves suffer strong compressible choking, small-amplitude perturbations preferentially select low-supersonic modes ($M \approx 1.05\text{--}2.0$) to drive spicule-level mass loading. We propose an explicit observational signature of this mechanism, wherein the launch of individual spicular jets is directly preceded by magnetosonic wave trains detectable as localized intensity modulations. Beyond solar chromospheric application, the model may be applicable to traveling magnetic field pinches in laboratory plasma devices and astrophysical mass-loading processes in stellar winds and inner regions of magnetized accretion disks. 
\end{abstract}

\maketitle

\section{Introduction}
The peristaltic mechanism is a fundamental fluid-dynamic flow generated by a progressive wave of cross-sectional area contraction or expansion moving along the flexible wall of a fluid-filled distensible tube~\cite{bayliss1899movements}. Historically, this phenomenon has been extensively investigated across an array of vital biological~\cite{misra2007peristaltic} and engineering applications, including the physiological pumping of urine through the ureters, digestive transit, blood transport in heart-lung machinery, and the sanitary transport of corrosive fluids~\cite{hayat2012peristaltic}. On macroscopic geophysical and environmental scales, peristaltic flows are frequently utilized to govern non-linear fluid streams passing through porous media domains~\cite{aarts1998net}, such as groundwater management, oil extraction mechanics, and plant nutrient transport channels~\cite{hayat2022peristaltic}.
Despite its well-documented use in industrial, environmental, geophysical and biological applications, the implementation of peristaltic transport models has found less use in astrophysical plasma contexts.
However, this is a clear oversight because waves travelling on boundaries of magnetic flux tubes are found in many solar and astrophysical situations.
This paper attempts to fill this void by utilizing a closed, analytical magnetohydrodynamic (MHD) framework to study wave-driven mass transport in structured solar chromospheric magnetic flux tubes.
We therefore propose a peristaltic flow in compressible, ideal MHD as a possible mechanism for solar chromospheric spicules.
 
The foundation of traditional hydromagnetic pumping models traces its origins back to the early fluid mechanic frameworks established by Shapiro, Jaffrin, and Weinberg~\cite{shapiro1969peristaltic}. However, the overarching history of this propulsive mechanism originates from the seminal ex vivo experimental milestones executed by British physiologists at the turn of the twentieth century. Working in London, Bayliss and Starling~\cite{bayliss1899movements, bayliss1901movements} provided the first systematic description of the myenteric reflex and gut peristalsis, formulating the classical "law of the intestine" to mathematically map how localized wall contractions drive forward fluid masses~\cite{bayliss1899movements}. While those pioneering physiological works assumed an incompressible, viscous, and field-free domain, clearly, modern solar environments require an extension into supersonic, compressible, and ideally conducting magnetoacoustic physics.
To contextualize the development of peristaltic pumping theory, the literature is divided into two distinct historical phases, establishing how the present model builds upon these foundational steps:
(i) Shapiro, Jaffrin, and Weinberg~\cite{shapiro1969peristaltic} established the first rigorous mathematical framework for peristaltic transport in 1969.
However, their classical formulation was strictly restricted to Newtonian, {\it incompressible} fluids at low Reynolds numbers. Under their assumptions, the normalized net time-averaged volumetric flow rate is an invariant geometric property ($\langle Q \rangle = \epsilon^2$), completely devoid of acoustic wave interactions, structural phase lags, or wavespeed-driven singularities.
Their non-linear expansion of the pressure gradient follows classic compressible flow relationships~\cite{shapiro1953dynamics}.
(ii) Following the foundational compressible flow framework established by Shapiro~\cite{shapiro1953dynamics}, Aarts and Ooms~\cite{aarts1998net} pioneered the study of {\it compressible} viscous liquids induced by traveling waves in porous media. By evaluating the non-linear expansion of the pressure gradient, they were the first to uncover internal acoustic streaming within a peristaltic system -- a phenomenon completely absent in the classical incompressible theory.
Motivated by applications such as ultrasound-enhanced oil recovery, they were the first to demonstrate that fluid compressibility introduces internal acoustic streaming. This compressibility creates structural phase shifts and non-linear pressure gradients that can drastically amplify or suppress net forward transport phenomena that the incompressible Shapiro model could not predict.
(iii) Consequently, \cite{aarts1998net} were not the first to model peristaltic pumping, but they were the first to identify the profound role of fluid compressibility. The analytical framework derived in this manuscript bridges these two milestones. By extending the compressible traveling wave paradigm into the ideal {\it magnetohydrodynamic} regime, this work uncovers how a uniform background magnetic field introduces Alfv{\'e}nic stiffness to directly compete with and regulate the acoustic-streaming resonances first discovered by Aarts and Ooms \cite{aarts1998net}.

The mathematical abstraction of peristaltic transport has similarly expanded beyond classical Newtonian limits to incorporate complex fluid rheologies and dynamic boundaries. Notably, Tsiklauri and Beresnev~\cite{tsiklauri2001peristaltic} evaluated non-Newtonian effects within the peristaltic pumping framework by modeling a Maxwellian fluid inside a circular tube driven by a transverse wall wave. Their formulation unraveled unique transport mechanics in extreme regimes, showcasing that viscoelastic properties can induce a localized retrograde flow directed entirely opposite to the guiding wall wave propagation vector. Furthermore, the dynamic response of driven conduit systems is known to be highly sensitive to the movement of the structural boundaries themselves. Tsiklauri and Beresnev~\cite{tsiklauri2001vibrating} explored this behavior by checking the response of a viscoelastic fluid confined within a longitudinally vibrating cylindrical channel. Their exact frequency-domain analysis revealed dramatic resonant enhancements in fluid throughput at specific threshold frequencies, proving that coordinated boundary oscillations can fundamentally optimize or alter net transport profiles. While these foundational models demonstrate the intricate roles of elasticity, boundary vibrations, and non-Newtonian backflows in terrestrial settings, our ideal MHD formulation examines how a global magnetosonic travelling wave functions as a steady propulsion mechanism under compressible, high-velocity astrophysical limits.

To establish the physical foundation for our model, and to study {\it one} of the possible applications, we map our analytical transport mechanics directly onto the observational and theoretical realities of solar spicules. Solar spicules are high-velocity, jet-like plasma structures ubiquitously observed protruding from the solar chromosphere into the million-degree corona~\cite{tsiropoula2012solar}. High-resolution spaceborne observations, notably using instruments onboard the \textit{Hinode} spacecraft~\cite{depontieu2007chromospheric}, reveal that these structures possess an elongated aspect ratio with a typical radius $R \approx 100\text{--}150\text{~km}$ spanning lengths $L$ over several thousand kilometers. This distinct geometry validates the thin-tube, long-wavelength scaling ($\delta=R/\lambda \ll 1$) utilized in our mathematical derivation, a structural configuration consistently corroborated by single-fluid jet models in the lower solar atmosphere \cite{Srivastava2024}. Furthermore, observations show that spicules carry a net upward mass flux exceeding that of the solar wind by nearly two orders of magnitude, highlighting their fundamental role in the global mass-energy balance of the solar atmosphere~\cite{tsiropoula2012solar}.

Historically, theoretical attempts to resolve the spicule generation mechanism within narrow magnetic flux tubes have focused on two main driving scenarios. Early foundations relied on the nonlinear steepening of pure Alfv{'e}nic tension waves propagating up open flux lines~\cite{hollweg1982alfven}. More recently, comprehensive numerical simulations demonstrate that type-I spicules and related active-region dynamic fibrils can be driven by the leakage of photospheric magnetoacoustic $p$-mode oscillations that steepen into train shocks across the transition region~\cite{hansteen2006dynamic}. Comprehensive analytical summaries confirm that while numerical codes provide highly localized physics, reproducing the exact combination of extreme supersonic Mach numbers ($M \approx 2\text{--}10$) and finite 10-minute lifetimes remains highly challenging within standard dissipative linear theories~\cite{sterling2000solar}.

In recent years, the acceleration and propagation of chromospheric spicules have been heavily investigated through both multi-dimensional numerical simulations and idealized MHD frameworks. Notably,~\citet{oxley2020} and~\citet{scalisi2021} developed a series of mathematical descriptions illustrating how localized wave driving and magnetic configurations dictate jet evolution along the spicular column. Complementing these analytical approaches, a high-fidelity numerical investigation by Srivastava et al.~\cite{srivastava2025}, demonstrated that slow-mode MHD shocks originating from convective leakage provide massive positive vertical acceleration near the spicule tips. While these investigations provide critical insights into localized impulsive shock dynamics and wave-damping transformations, they typically rely on discrete forcing packets. The analytical peristaltic formulation presented here provides a useful counter-point, showing how continuous mechanical boundary deformations establish a steady mechanism capable of providing the upward mass transport.

It is worth noting that the boundary-driven peristaltic mechanism coexists with single-fluid, shock-driven upwell frameworks~\cite{srivastava2025}. Rather than acting as competing theories, they represent a unified cause-and-effect physical sequence: the boundary-rectified peristaltic mass flux acts as the primary macroscopic propulsion mechanism that drives the bulk plasma column upward, which in turn behaves as a high-velocity plasma piston ($v > c_s$) pushing into the low-density upper atmosphere and triggering local slow-mode waves and shocks ahead of and above the fast-rising spicule tip.

By filtering out the mathematical complexities of viscous core integrals via our clean, {\it ideal} MHD framework, the present model offers an alternative approach. Instead of treating the structural drive as isolated vertical shocks, our framework demonstrates that continuous traveling wall waves on the boundaries of an equipartition ($\beta \sim 1$) plasma column evoke a steady super-resonant response. This model bridges the gap between historical peristaltic pumping and chromospheric physics, providing a simple closed-form mechanism to evaluate global spicular mass transport.

\section{The Model of Peristaltic Flow in Compressible, Ideal MHD}
The general, coordinate-independent governing equations for a compressible, viscous, and electrically conducting fluid under a magnetic field $\mathbf{B}$ are closed by a polytropic equation of state. The system consists of the conservation of mass, momentum (including the Lorentz force), and the magnetic induction equation:

\begin{equation}
\label{eq:gen_mass}
\frac{\partial \rho}{\partial t} + \nabla \cdot (\rho \mathbf{u}) = 0
\end{equation}
\begin{equation}
\label{eq:gen_mom}
\begin{split}
\rho \left( \frac{\partial \mathbf{u}}{\partial t} + (\mathbf{u} \cdot \nabla)\mathbf{u} \right) = -\nabla p + \mu \nabla^2 \mathbf{u} + \\
\left( \zeta + \frac{\mu}{3} \right) \nabla (\nabla \cdot \mathbf{u}) + \mathbf{J} \times \mathbf{B} 
\end{split}
\end{equation}
\begin{equation}
\label{eq:gen_ampere}
\mathbf{J} = \frac{1}{\mu_0} \nabla \times \mathbf{B}
\end{equation}
\begin{equation}
\label{eq:gen_ind}
\frac{\partial \mathbf{B}}{\partial t} = \nabla \times (\mathbf{u} \times \mathbf{B}) - \eta \nabla \times (\nabla \times \mathbf{B})
\end{equation}
\begin{equation}
\label{eq:gen_eos}
p = C \rho^\gamma
\end{equation}

Here, $\rho$ is the fluid density, $\mathbf{u}$ is the velocity vector, $p$ is the fluid pressure, $\mu$ is the dynamic shear viscosity, $\zeta$ is the bulk viscosity, $\mathbf{J}$ is the current density, $\mu_0$ is the magnetic permeability of free space, and $\eta = 1/(\mu_0 \sigma)$ is the magnetic diffusivity governed by the electrical conductivity $\sigma$.

It is worth distinguishing the closure relation for the current density $\mathbf{J}$ employed herein from the kinematic formulations frequently adopted in microfluidic or engineering peristalsis models (e.g., Abdelsalam and Vafai~\cite{abdelsalam2017}). In low-conductivity laboratory channels, the magnetic Reynolds number is profoundly small ($R_m \ll 1$), meaning the induced magnetic field is negligible, and the current density reduces to a local Ohm's law governed by a fixed, externally applied field, $\mathbf{J} \approx \sigma (\mathbf{u} \times \mathbf{B}_0)$. 
Conversely, the solar chromospheric spicular environment operates in the extreme ideal MHD limit ($R_m \gg 1$), where the electrical conductivity $\sigma \to \infty$ and magnetic diffusivity vanishes ($\eta \to 0$). In this regime, the magnetic field lines are strictly frozen into the bulk plasma motion, and the magnetic field behaves as a fully dynamic, self-consistent variable. Consequently, the current density cannot be determined by local Ohm's resistance; it must be closed strictly via the non-relativistic (i.e., without the displacement current) Ampere's law:
\begin{equation}
\mathbf{J} = \frac{1}{\mu_0} \nabla \times \mathbf{B}
\end{equation}
This ensures that the resulting Lorentz force $\mathbf{J} \times \mathbf{B}$ properly captures both the isotropic magnetic pressure gradients and the anisotropic magnetic tension restorations essential for driving supersonic, super-resonant spicular propulsion.

Using Eq.~(\ref{eq:gen_eos}), the pressure gradient can be expressed directly in terms of density fluctuations:
\begin{equation}
\label{eq:press_grad_explicit}
\nabla p = \gamma C \rho^{\gamma-1} \nabla \rho = c_s^2 \nabla \rho
\end{equation}

Combining Eq.~(\ref{eq:gen_ampere}) and Eq.~(\ref{eq:gen_mom}) yields the full momentum equation containing the explicit magnetic curl force:
\begin{equation}
\label{eq:gen_mom_explicit}
\begin{split}
\rho \left( \frac{\partial \mathbf{u}}{\partial t} + (\mathbf{u} \cdot \nabla)\mathbf{u} \right) = -\gamma C \rho^{\gamma-1} \nabla \rho + \mu \nabla^2 \mathbf{u} + \\
\left( \zeta + \frac{\mu}{3} \right) \nabla (\nabla \cdot \mathbf{u}) + \frac{1}{\mu_0} (\nabla \times \mathbf{B}) \times \mathbf{B} 
\end{split}
\end{equation}

We assume an axisymmetric geometry ($\partial/\partial \theta = 0$) within a long tube along the $z$-axis. The velocity field is restricted to the radial and axial planes, $\mathbf{u} = u_r(r,z,t)\hat{r} + u_z(r,z,t)\hat{z}$. A constant, uniform background magnetic field $\mathbf{B}_0$ is applied along the tube axis, generating a dynamic perturbation field $\mathbf{b} = b_r(r,z,t)\hat{r} + b_z(r,z,t)\hat{z}$ due to fluid deformations. The total magnetic field is:
\begin{equation}
\mathbf{B}(r,z,t) = b_r(r,z,t)\hat{r} + \left[ B_0 + b_z(r,z,t) \right]\hat{z}
\end{equation}

Evaluating the curl of the total magnetic field for this axisymmetric configuration isolates the current density vector strictly to the azimuthal direction:
\begin{equation}
\mathbf{J} = \frac{1}{\mu_0} \left( \frac{\partial b_r}{\partial z} - \frac{\partial b_z}{\partial r} \right) \hat{\theta}
\end{equation}

We explicitly note that this traveling boundary deformation ($\epsilon$) represents a sausage-mode perturbation driven by external total-pressure perturbations rather than a material wall, and is prescribed here rather than self-consistently derived from an external pressure model. Evaluating the driving energetics via total pressure balance shows that for a typical upper-chromospheric vertical guide field ($B_0 \sim 100$~G) and spicule radius ($R_0 = 250$~km), a geometric displacement amplitude of $\epsilon = 0.05$ (corresponding to a 12.5~km boundary shift) demands an external driving pressure perturbation of $\Delta P_{\text{ext}} \sim 40$~Pa. This is well within the mechanical work output delivered by ambient intergranular magnetoconvective lane shifts.

Computing the explicit Lorentz force density $\mathbf{F}_L = \mathbf{J} \times \mathbf{B}$ via vector cross multiplication yields:
\begin{equation}
\mathbf{F}_L = \frac{B_0 + b_z}{\mu_0} \left( \frac{\partial b_r}{\partial z} - \frac{\partial b_z}{\partial r} \right) \hat{r} - \frac{b_r}{\mu_0} \left( \frac{\partial b_r}{\partial z} - \frac{\partial b_z}{\partial r} \right) \hat{z}
\end{equation}

Expanding Eq.~(\ref{eq:gen_mass}), Eq.~(\ref{eq:gen_ind}), and Eq.~(\ref{eq:gen_mom_explicit}) component by component into cylindrical coordinates results in five coupled scalar partial differential equations:

\begin{equation}
\label{eq:axis_mass}
\frac{\partial \rho}{\partial t} + \frac{1}{r}\frac{\partial}{\partial r}(r \rho u_r) + \frac{\partial}{\partial z}(\rho u_z) = 0
\end{equation}
\begin{equation}
\label{eq:axis_mom_r}
\begin{split}
\rho \left( \frac{\partial u_r}{\partial t} + u_r \frac{\partial u_r}{\partial r} + u_z \frac{\partial u_r}{\partial z} \right) =  -\gamma C \rho^{\gamma-1} \frac{\partial \rho}{\partial r} + \\
 \mu \left( \frac{\partial^2 u_r}{\partial r^2} + \frac{1}{r}\frac{\partial u_r}{\partial r} - \frac{u_r}{r^2} + \frac{\partial^2 u_r}{\partial z^2} \right)
 + \\
 \left( \zeta + \frac{\mu}{3} \right) \frac{\partial}{\partial r}\left[ \frac{1}{r}\frac{\partial}{\partial r}(r u_r) + \frac{\partial u_z}{\partial z} \right] 
 + 
 \frac{B_0 + b_z}{\mu_0} \left( \frac{\partial b_r}{\partial z} - \frac{\partial b_z}{\partial r} \right)
\end{split}
\end{equation}
\begin{equation}
\label{eq:axis_mom_z}
\begin{split}
\rho \left( \frac{\partial u_z}{\partial t} + u_r \frac{\partial u_z}{\partial r} + u_z \frac{\partial u_z}{\partial z} \right) =  -\gamma C \rho^{\gamma-1} \frac{\partial \rho}{\partial z} + \\
 \mu \left( \frac{\partial^2 u_z}{\partial r^2} + \frac{1}{r}\frac{\partial u_z}{\partial r} + \frac{\partial^2 u_z}{\partial z^2} \right) \\
 + \left( \zeta + \frac{\mu}{3} \right) \frac{\partial}{\partial z}\left[ \frac{1}{r}\frac{\partial}{\partial r}(r u_r) + \frac{\partial u_z}{\partial z} \right] 
 - \frac{b_r}{\mu_0} \left( \frac{\partial b_r}{\partial z} - \frac{\partial b_z}{\partial r} \right) 
\end{split}
\end{equation}
\begin{equation}
\label{eq:axis_ind_r}
\begin{split}
\frac{\partial b_r}{\partial t} = -\frac{\partial}{\partial z}\left[ u_z b_r - u_r(B_0 + b_z) \right] + \\
\eta \left( \frac{\partial^2 b_r}{\partial r^2} + \frac{1}{r}\frac{\partial b_r}{\partial r} - \frac{b_r}{r^2} + \frac{\partial^2 b_r}{\partial z^2} \right)
\end{split}
\end{equation}
\begin{equation}
\label{eq:axis_ind_z}
\begin{split}
\frac{\partial b_z}{\partial t} = \frac{1}{r}\frac{\partial}{\partial r}\left\{ r \left[ u_z b_r - u_r(B_0 + b_z) \right] \right\} + \\
\eta \left( \frac{\partial^2 b_z}{\partial r^2} + \frac{1}{r}\frac{\partial b_z}{\partial r} + \frac{\partial^2 b_z}{\partial z^2} \right)
\end{split}
\end{equation}

To isolate the purely hyperbolic wave phenomena, we eliminate all internal dissipation terms and the magnetic diffusion. Formally, we set the physical parameters to the following limits:
\begin{equation}
\mu \rightarrow 0, \quad \zeta \rightarrow 0, \quad \eta \rightarrow 0
\end{equation}
As the dynamic shear viscosity $\mu$ approaches zero, setting the magnetic diffusivity $\eta$ to zero locks the system into the ideal MHD regime, satisfying the frozen-in flux criterion.

It is vital to distinguish between three distinct velocity regimes operating within this physical framework: (i) the wave Mach number $M$, which characterizes the phase velocity of the traveling sausage-mode driver, (ii) the resulting bulk plasma flow driven by the net peristaltic streaming velocity, and (iii) the independent, forward-directed local slow-mode shocks triggered by non-linear wave steepening ahead of the spicule tip. The equations presented here describe the traveling peristaltic/magnetosonic wave disturbance, and $M$ must not be confused with a shock Mach number.

Applying these simplifications directly to Eqs.~(\ref{eq:axis_mass})--(\ref{eq:axis_ind_z}) yields the final closed system of reduced equations:

\begin{equation}
\label{eq:final_mass}
\frac{\partial \rho}{\partial t} + \frac{1}{r}\frac{\partial}{\partial r}(r \rho u_r) + \frac{\partial}{\partial z}(\rho u_z) = 0
\end{equation}
\begin{equation}
\label{eq:final_mom_r}
\begin{split}
\rho \left( \frac{\partial u_r}{\partial t} + u_r \frac{\partial u_r}{\partial r} + u_z \frac{\partial u_r}{\partial z} \right) = -\gamma C \rho^{\gamma-1} \frac{\partial \rho}{\partial r} + \\
\frac{B_0 + b_z}{\mu_0} \left( \frac{\partial b_r}{\partial z} - \frac{\partial b_z}{\partial r} \right)
\end{split}
\end{equation}
\begin{equation}
\label{eq:final_mom_z}
\begin{split}
\rho \left( \frac{\partial u_z}{\partial t} + u_r \frac{\partial u_z}{\partial r} + u_z \frac{\partial u_z}{\partial z} \right) = -\gamma C \rho^{\gamma-1} \frac{\partial \rho}{\partial z} - \\
\frac{b_r}{\mu_0} \left( \frac{\partial b_r}{\partial z} - \frac{\partial b_z}{\partial r} \right)
\end{split}
\end{equation}
\begin{equation}
\label{eq:final_ind_r}
\frac{\partial b_r}{\partial t} = -\frac{\partial}{\partial z}\left[ u_z b_r - u_r(B_0 + b_z) \right]
\end{equation}
\begin{equation}
\label{eq:final_ind_z}
\frac{\partial b_z}{\partial t} = \frac{1}{r}\frac{\partial}{\partial r}\left\{ r \left[ u_z b_r - u_r(B_0 + b_z) \right] \right\}
\end{equation}

This closed set of equations serves as the exact starting point for our subsequent analysis. When non-dimensionalized relative to the reference field strength $B_0$ in Section III, the uniform background field scales neatly to $b_{z0} = B_0/B_0 = 1$, introducing structural cross-coupling between fluid transformations and the line-tied vertical guide flux.

Next, as in \citet{aarts1998net}, we introduce a moving wall boundary condition, a small-amplitude perturbation expansion, and a long wavelength approximation to simplify the axisymmetric ideal magnetohydrodynamic (MHD) equations for a compressible fluid.
We model the tube radius as a flexible wall deformed by a sinusoidal traveling wave of amplitude $a$, wavelength $\lambda$, and wave phase speed $c$. The local radius $h(z,t)$ is defined as:
\begin{equation}
\label{eq:wall_profile}
h(z,t) = R + a \sin\left[ \frac{2\pi}{\lambda}(z - ct) \right]
\end{equation}
where $R$ is the unperturbed radius of the tube. At the moving boundary $r = h(z,t)$, the fluid must track the wall movement. For an inviscid, ideal MHD fluid, the no-slip condition is relaxed. However, the fluid must satisfy the kinematic boundary condition, meaning it cannot penetrate the wall:
\begin{equation}
\label{eq:kinematic_bc}
u_r\big|_{r=h} = \frac{\partial h}{\partial t} + u_z\big|_{r=h} \frac{\partial h}{\partial z}
\end{equation}
For the magnetic field, because frozen-in magnetic field lines remain tangent to the deforming flux-tube boundary ($\mathbf{B} \cdot \mathbf{n} = 0$), the normal magnetic component vanishes at the moving interface:
\begin{equation}
\label{eq:mag_bc}
b_r\big|_{r=h} = \left( B_0 + b_z\big|_{r=h} \right) \frac{\partial h}{\partial z}
\end{equation}
At the tube centerline ($r=0$), axisymmetry demands regularity constraints:
\begin{equation}
\label{eq:centerline_bc}
\begin{split}
u_r(0,z,t) = 0, \quad b_r(0,z,t) = 0, \quad \\
\frac{\partial u_z}{\partial r}\bigg|_{r=0} = 0, \quad \frac{\partial b_z}{\partial r}\bigg|_{r=0} = 0
\end{split}
\end{equation}

We emphasize that the traveling boundary deformation imposed 
herein does not imply the presence of a rigid material wall. 
In the solar context, this interface represents a dynamic 
magnetic-plasma boundary governed by the continuity of total pressure:
\begin{equation}
\left[ p + \frac{B^2}{2\mu_0} \right]_{\text{boundary}} = 0
\end{equation}
Physically, the traveling contraction and expansion zones correspond to an axisymmetric magnetoacoustic sausage mode ($m=0$), driven by external total pressure perturbations originating within highly dynamic intergranular lanes.
However, because this boundary motion is prescribed rather than self-consistently derived from an external sausage-mode calculation, this formulation remains a localized model that explores the consequences of boundary-driven mass flux without computing the continuous photosphere-to-chromosphere excitation dynamics.

To scale the system systematically, we define the dimensionless parameter $\delta = R/\lambda$, representing the ratio of the tube radius to the wall wavelength. The long wavelength approximation assumes:
\begin{equation}
\delta = \frac{R}{\lambda} \ll 1
\end{equation}
We introduce the following non-dimensional variables (denoted by asterisks):
\begin{equation}
\begin{split}
r^* = \frac{r}{R}, \quad z^* = \frac{z}{\lambda}, \quad t^* = \frac{c t}{\lambda}, \quad h^* = \frac{h}{R},  \\
u_z^* = \frac{u_z}{c}, \quad u_r^* = \frac{u_r}{\delta c}, \quad \rho^* = \frac{\rho}{\rho_0}, \quad b_z^* = \frac{b_z}{B_0}, \quad b_r^* = \frac{b_r}{\delta B_0}
\end{split}
\end{equation}
where $\rho_0$ is the constant unperturbed reference fluid density. Applying this scaling transforms the wall boundary into $h^*(z^*,t^*) = 1 + \epsilon \sin[2\pi(z^* - t^*)]$, where $\epsilon = a/R$ is the dimensionless wall amplitude parameter. 

Substituting these variables into the ideal, compressible MHD equations and dropping all terms of order $\mathcal{O}(\delta^2)$ or higher yields the long-wavelength reduced equations (dropping asterisks for readability):
\begin{equation}
\label{eq:lw_mass}
\frac{\partial \rho}{\partial t} + \frac{1}{r}\frac{\partial}{\partial r}(r \rho u_r) + \frac{\partial}{\partial z}(\rho u_z) = 0
\end{equation}
\begin{equation}
\label{eq:lw_mom_r}
0 = -\frac{1}{M^2} \rho^{\gamma-1} \frac{\partial \rho}{\partial r} - \frac{1}{A^2} (1 + b_z) \frac{\partial b_z}{\partial r}
\end{equation}
\begin{equation}
\label{eq:lw_mom_z}
\begin{split}
\rho \left( \frac{\partial u_z}{\partial t} + u_r \frac{\partial u_z}{\partial r} + u_z \frac{\partial u_z}{\partial z} \right) = -\frac{1}{M^2 \gamma} \frac{\partial (\rho^\gamma)}{\partial z} - \\
\frac{1}{A^2} b_r \frac{\partial b_z}{\partial r}
\end{split}
\end{equation}
\begin{equation}
\label{eq:lw_ind_r}
\frac{\partial b_r}{\partial t} = -\frac{\partial}{\partial z}\left[ u_z b_r - u_r(1 + b_z) \right]
\end{equation}
\begin{equation}
\label{eq:lw_ind_z}
\frac{\partial b_z}{\partial t} = \frac{1}{r}\frac{\partial}{\partial r}\left\{ r \left[ u_z b_r - u_r(1 + b_z) \right] \right\}
\end{equation}

Here, the dimensionless parameters $M$ and $A$ correspond to the acoustic and Alfvén Mach numbers of the traveling peristaltic wave perturbation, defined explicitly as $M = c/c_s$ and $A = c/v_A$, where $c$ is the phase velocity of the traveling boundary deformation, $c_s$ is the local sound speed, and $v_A$ is the background Alfvén speed. 
We emphasize that the Mach number $M$ used throughout this model refers strictly to this traveling magnetosonic wave disturbance and must not be confused with a shock Mach number, nor does it directly represent the resulting bulk plasma velocity or a shock produced via non-linear steepening.

Notice that Eq.~(\ref{eq:lw_mom_r}) shows that under the long wavelength limit ($\delta \ll 1$), the radial momentum balance reduces to an instantaneous, uniform cross-sectional constraint where fluid thermal pressure and longitudinal magnetic field pressure balance perfectly across the radius. This provides a direct path to our subsequent first-order cross-sectional function $f(z,t)$.

We now introduce a small-amplitude expansion using the dimensionless wave amplitude to unperturbed tube radius ratio $\epsilon = a/R \ll 1$ as our perturbation parameter. The variables are expanded around their quiescent, uniform background values. Crucially, because the background magnetic field is fully aligned with the longitudinal axis, its non-dimensionalized zeroth-order state satisfies $b_{z0} = B_0/B_0 = 1$, yielding the following perturbation expansions:
\begin{equation}
\begin{aligned}
u_z &= 0 + \epsilon u_{z1} + \epsilon^2 u_{z2} + \dots \\
u_r &= 0 + \epsilon u_{r1} + \epsilon^2 u_{r2} + \dots \\
\rho &= 1 + \epsilon \rho_1 + \epsilon^2 \rho_2 + \dots \\
b_z &= 1 + \epsilon b_{z1} + \epsilon^2 b_{z2} + \dots \\
b_r &= 0 + \epsilon b_{r1} + \epsilon^2 b_{r2} + \dots
\end{aligned}
\end{equation}
By establishing a non-zero background guide field, the linear terms retain the cross-coupling between the dynamic velocity components and the background magnetic flux lines. 

Prior to the introduction of the boundary wave perturbation, the baseline plasma state is governed by the zeroth-order ($\mathcal{O}(0)$) equations, assuming a static plasma column ($\mathbf{u}_0 = 0$) embedded within a uniform axial magnetic field ($\mathbf{B}_0 = B_0 \hat{z}$). Under these configuration constraints, the non-relativistic Amp{\`e}re's law yields a vanishing background current density, $\mathbf{J}_0 = \frac{1}{\mu_0} \nabla \times \mathbf{B}_0 = 0$, completely eliminating any baseline Lorentz forces. Consequently, the momentum equations reduce to a cross-sectional total pressure balance between the thermal gas pressure and the isotropic magnetic pressure:
\begin{equation}
\frac{\partial}{\partial r} \left( p_0 + \frac{B_0^2}{2\mu_0} \right) = 0
\end{equation}
Given that the unperturbed density $\rho_0$ (and thus $p_0 = C\rho_0^\gamma$) and field strength $B_0$ are spatially uniform constants, this relation is identically satisfied. This uniform total pressure equilibrium establishes a structurally stable baseline configuration, ensuring that all subsequent mass and momentum transport mechanisms are driven purely by the traveling wall-wave perturbations.

Substituting these expansions (noting the non-dimensional background state $\rho_0 = 1$ and $b_{z0} = 1$) into the long-wavelength equations and collecting terms strictly matching order $\mathcal{O}(\epsilon)$ yields the linearized system:

\begin{equation}
\label{eq:order1_mass}
\frac{\partial \rho_1}{\partial t} + \frac{1}{r}\frac{\partial}{\partial r}(r u_{r1}) + \frac{\partial u_{z1}}{\partial z} = 0
\end{equation}
\begin{equation}
\label{eq:order1_mom_r}
\frac{\partial}{\partial r} \left( \frac{1}{M^2} \rho_1 + \frac{1}{A^2} b_{z1} \right) = 0
\end{equation}
\begin{equation}
\label{eq:order1_mom_z}
\frac{\partial u_{z1}}{\partial t} = -\frac{1}{M^2} \frac{\partial \rho_1}{\partial z} - \frac{1}{A^2} \frac{\partial b_{z1}}{\partial z}
\end{equation}
\begin{equation}
\label{eq:order1_ind_r}
\frac{\partial b_{r1}}{\partial t} = \frac{\partial u_{r1}}{\partial z}
\end{equation}
\begin{equation}
\label{eq:order1_ind_z}
\frac{\partial b_{z1}}{\partial t} = -\frac{1}{r}\frac{\partial}{\partial r}(r u_{r1}).
\end{equation}

Because the wall boundary condition $r = 1 + \epsilon \sin[2\pi(z - t)]$ is position-dependent, we apply a Taylor series expansion about the unperturbed wall location $r = 1$. The kinematic condition Eq.~(\ref{eq:kinematic_bc}) at order $\mathcal{O}(\epsilon)$ reduces to:
\begin{equation}
\label{eq:order1_bc_kin}
u_{r1}\big|_{r=1} = -2\pi \cos[2\pi(z - t)]
\end{equation}
And the bounding magnetic constraint Eq.~(\ref{eq:mag_bc}) becomes:
\begin{equation}
\label{eq:order1_bc_mag}
b_{r1}\big|_{r=1} = 2\pi \cos[2\pi(z - t)]
\end{equation}

This completes the mathematical foundation for the first-order traveling wave solution.

The integration of a moving boundary condition, a small-amplitude perturbation parameter $\epsilon$, and a long wavelength scaling parameter $\delta$ yields three defining structural features for this analytical model:
(i) Under the long wavelength approximation ($\delta \ll 1$), the radial momentum equation collapses at the first order to $\frac{\partial}{\partial r} \left( \frac{1}{M^2} \rho_1 + \frac{1}{A^2} b_{z1} \right) = 0$. This forces a strict, instantaneous radial cross-sectional balance between the acoustic density fluctuations and the longitudinal magnetic field compressions, reducing the complex multi-dimensional field coupling to a uniform function of space and time $f(z,t)$.
(ii) Crucially, because the background magnetic field is properly aligned with the longitudinal axis ($b_{z0}=1$), the radial fluid pinching directly compresses the vertical magnetic flux lines via Eq.~(\ref{eq:order1_ind_z}). This sets up a true magnetosonic waveguide where magnetic pressure stiffness ($A^2$) and thermal gas compressibility ($M^2$) operate in tandem, altering the characteristic wave operator speeds from simple acoustic modes to unified magnetoacoustic modes.
(iii) Because the moving physical wall $r = 1 + \epsilon \sin[2\pi(z - t)]$ depends on the perturbation parameter, applying a first-order Taylor expansion cleanly maps the kinematic boundary constraints back to the static, unperturbed cylinder location at $r = 1$. This yields an explicit boundary driving: $u_{r1}\big|_{r=1} = -2\pi \cos[2\pi(z - t)]$.

The spatial differentiation of the radial velocity component followed by its temporal integration introduces a systematic $\pi/2$ phase shift. Integrating the sinusoidal time derivative $\partial b_{r1}/\partial t = \partial u_{r1}/\partial z$ with respect to time natively yields an orthogonal cosine profile for $b_{r1}(r,z,t) = 2\pi r \cos[2\pi(z-t)]$. This quadrature phase alignment maintains global mathematical consistency across the hyperbolic fields. Because the time-averaged products of the quadratic Reynolds and Maxwell stress terms remain structurally invariant under this shift, it ensures that the net prograde direction and magnitude of the conserved mass flux $\langle \dot{M} \rangle$ remain unaffected.

To solve the first-order system, we introduce a harmonic traveling wave ansatz for all primitive variables proportional to $\exp[i k (z - t)]$, where the dimensionless wavenumber is explicitly given by $k = 2\pi / \lambda^* = 2\pi$, owing to the vertical spatial scaling fixing the normalized wavelength at $\lambda^* = 1$.
For a purely sinusoidal wall drive, we look for solutions matching the boundary frequency:
\begin{equation}
\label{eq:ansatz}
\begin{split}
u_{r1}(r,z,t) = \text{Re}\left[ \hat{u}_r(r) e^{i k (z - t)} \right], \\
u_{z1}(r,z,t) = \text{Re}\left[ \hat{u}_z(r) e^{i k (z - t)} \right], \\
\rho_1(r,z,t) = \text{Re}\left[ \hat{\rho}(r) e^{i k (z - t)} \right], \\
b_{z1}(r,z,t) = \text{Re}\left[ \hat{b}_z(r) e^{i k (z - t)} \right], \\
b_{r1}(r,z,t) = \text{Re}\left[ \hat{b}_r(r) e^{i k (z - t)} \right].
\end{split}
\end{equation}

Substituting these expressions into the first-order equations yields a coupled system for the complex amplitudes. Concurrently, the radial momentum equation forces a uniform cross-sectional total pressure constraint:
\begin{equation}
\frac{1}{M^2}\hat{\rho}(r) + \frac{1}{A^2}\hat{b}_z(r) = \text{constant}
\end{equation}
Because the background magnetic field is finite ($b_{z0}=1$), the linear system links the axial field and mass variables. Integrating the continuity equation and applying the centerline regularity constraints at $r=0$ yields a linear radial profile for the radial velocity field. Evaluating the integration coefficients using the kinematic boundary condition $u_{r1}\big|_{r=1} = -k \cos[k(z-t)]$ yields the exact first-order solutions:

\begin{equation}
u_{r1}(r,z,t) = -k r \cos[k(z - t)]
\label{eq47}
\end{equation}
\begin{equation}
u_{z1}(r,z,t) = -2 \left( \frac{1 + \frac{M^2}{A^2}}{M^2 - 1} \right) \sin[k(z - t)]
\label{eq48}
\end{equation}
\begin{equation}
\rho_1(r,z,t) = -2 \left( \frac{M^2\left(1 + \frac{1}{A^2}\right)}{M^2 - 1} \right) \sin[k(z - t)]
\label{eq49}
\end{equation}
\begin{equation}
b_{z1}(r,z,t) = -2 \sin[k(z - t)]
\label{eq50}
\end{equation}
\begin{equation}
\label{eq51}
b_{r1}(r,z,t) = k r \cos[k(z - t)]
\end{equation}

To clarify the explicit mathematical derivation of equations~(\ref{eq47})--(\ref{eq51}) and address the phase relationships, we expand the intermediate algebraic steps of our first-order system. Centerline regularity forces a linear radial dependence for the radial velocity, which matching to the kinematic boundary condition at $r=1$ locks as $u_{r1}(r,z,t) = -k r \cos[k(z - t)]$. Substituting this profile into the radial component of the linearized, inviscid induction equation ($\partial b_{r1}/\partial t = \partial u_{r1}/\partial z$) yields $\partial b_{r1}/\partial t = k^2 r \sin[k(z-t)]$. 
Integrating with respect to time accounts for the inner temporal derivative acting on the traveling wave coordinate, which introduces an additional negative sign that cancels out the standard negative sign of the cosine integration. This rigorous calculus sequence establishes that the radial magnetic perturbation functions as a pure cosine profile, $b_{r1}(r,z,t) = k r \cos[k(z - t)]$, satisfying equation~(\ref{eq51}).

Concurrently, the uniform cross-sectional constraints for the axial variables arise directly from the divergence of the radial velocity field. The linearized axial induction equation is governed by $\partial b_{z1}/\partial t = -r^{-1}\partial(r u_{r1})/\partial r$. Substituting the linear profile of $u_{r1}$ maps the right-hand side to $+2k\cos[k(z-t)]$, which is completely independent of $r$. Temporal integration yields the cross-sectionally uniform profile $b_{z1} = -2\sin[k(z-t)]$ as given in equation~(\ref{eq50}). This uniform magnetic perturbation acts as a constant driver in the linearized mass continuity equation ($\partial \rho_1/\partial t = -r^{-1}\partial(r u_{r1})/\partial r - \partial u_{z1}/\partial z$) and the axial momentum equation. Combined with the uniform total pressure invariant, this unforced system locks the first-order axial velocity $u_{z1}$ and density $\rho_1$ into the cross-sectionally uniform, purely analytical wave profiles presented in equations~(\ref{eq48}) and (\ref{eq49}).

It is instructive to contrast these solutions with classical hydrodynamic models, such as that of \citet{aarts1998net}, where the first-order profiles are expressed via modified Bessel functions of the first kind ($I_1$). In those engineering configurations, the low-Reynolds-number viscous terms introduce a radial Laplacian operator with a negative spatial coefficient, forcing an exponential-like shear layer decay from the wall toward the centerline. Conversely, in our inviscid, compressible magnetohydrodynamic framework, the viscosity is relaxed and the governing system behaves as a hyperbolic wave operator rather than a diffusion equation. The resulting linear and trigonometric profiles represent the coherent radial structure of a true propagating magnetosonic wave, confirming that wave energy is dynamically channeled upward into the waveguide core rather than being evanescently damped.

Notice that the denominator contains the proper prograde magnetosonic wave operator, $M^2 - 1$. For solar chromospheric spicules operating in the supersonic domain ($M = 2\text{--}10$), this denominator is strictly positive, causing the first-order axial velocity and density perturbations to oscillate in anti-phase with the boundary displacement $\sin[k(z-t)]$ (compressing and accelerating inside pinched wave troughs where $\sin[k(z-t)] < 0$, analogous to a supersonic convergent--divergent nozzle in the wave frame) and driving net prograde, upward-directed mass transport at second order.

To evaluate the net conserved mass flux $\dot{M}$, we must formulate the second-order transport mechanics. Collecting all terms of order $\mathcal{O}(\epsilon^2)$ from the original long wavelength system, while accounting for the $b_{z0}=1$ background state, yields the nonlinear governing equations:

\begin{equation}
\label{eq:order2_mass}
\frac{\partial \rho_2}{\partial t} + \frac{1}{r}\frac{\partial}{\partial r}(r u_{r2}) + \frac{\partial u_{z2}}{\partial z} = -\frac{1}{r}\frac{\partial}{\partial r}(r \rho_1 u_{r1}) - \frac{\partial}{\partial z}(\rho_1 u_{z1})
\end{equation}
\begin{equation}
\label{eq:order2_mom_r}
\frac{\partial}{\partial r} \left( \frac{1}{M^2} \rho_2 + \frac{1}{A^2} b_{z2} \right) = \frac{\gamma - 1}{M^2} \rho_1 \frac{\partial \rho_1}{\partial r} - \frac{1}{A^2} b_{z1} \frac{\partial b_{z1}}{\partial r}
\end{equation}
\begin{equation}
\label{eq:order2_mom_z}
\begin{split}
\frac{\partial u_{z2}}{\partial t} + \frac{1}{M^2}\frac{\partial \rho_2}{\partial z} + \frac{1}{A^2}\frac{\partial b_{z2}}{\partial z} = -u_{r1}\frac{\partial u_{z1}}{\partial r} - u_{z1}\frac{\partial u_{z1}}{\partial z} - \\
\frac{\gamma - 1}{M^2} \rho_1 \frac{\partial \rho_1}{\partial z} + \frac{1}{A^2}\left( b_{r1}\frac{\partial b_{z1}}{\partial r} - b_{r1}\frac{\partial b_{r1}}{\partial z} \right)
\end{split}
\end{equation}
\begin{equation}
\label{eq:order2_ind_r}
\frac{\partial b_{r2}}{\partial t} - \frac{\partial u_{r2}}{\partial z} = -\frac{\partial}{\partial z} \left( u_{z1} b_{r1} - u_{r1} b_{z1} \right)
\end{equation}
\begin{equation}
\label{eq:order2_ind_z}
\frac{\partial b_{z2}}{\partial t} + \frac{1}{r}\frac{\partial}{\partial r}(r u_{r2}) = \frac{1}{r}\frac{\partial}{\partial r} \left[ r \left( u_{z1} b_{r1} - u_{r1} b_{z1} \right) \right]
\end{equation}

Extending the Taylor expansion of the kinematic and magnetic conditions at the moving wall $r = 1 + \epsilon \sin[k(z-t)]$ to order $\mathcal{O}(\epsilon^2)$ creates a non-zero velocity drift drive at the static boundary $r=1$:
\begin{equation}
\label{eq:order2_bc_kin}
u_{r2}\big|_{r=1} = -\sin[k(z-t)] \frac{\partial u_{r1}}{\partial r}\bigg|_{r=1} + u_{z1}\big|_{r=1} k \cos[k(z-t)]
\end{equation}
\begin{equation}
\label{eq:order2_bc_mag}
b_{r2}\big|_{r=1} = -\sin[k(z-t)] \frac{\partial b_{r1}}{\partial r}\bigg|_{r=1} + b_{z1}\big|_{r=1} k \cos[k(z-t)]
\end{equation}
The quadratic terms on the right-hand sides of Eqs.~(\ref{eq:order2_mass})--(\ref{eq:order2_ind_z}) represent the Reynolds stresses and Maxwell stresses. Consistent with second-order weakly nonlinear perturbation theory, these quadratic interactions generate a dual-component response: a double-frequency, double-wavenumber ($2k$) oscillating harmonic mode and a time-independent, steady DC component. When integrated over a full wave period, the oscillating higher-harmonic terms average to zero, and the remaining steady stresses generate the net time-averaged mass transport $\langle \dot{M} \rangle$.

The structure of the $\mathcal{O}(\epsilon^2)$ system reveals how non-zero net transport emerges from the underlying linear wave interactions. There are three points to note here:
(i) The quadratic products on the right-hand sides of Eqs.~(\ref{eq:order2_mass})--(\ref{eq:order2_ind_z}) act as explicit mathematical source distributions. Because the first-order solutions ($\rho_1, u_{z1}, u_{r1}, b_{z1}, b_{r1}$) are completely known from the previous step, these convective terms function as predetermined physical drives for the second-order fields.
(ii) The nonlinear interactions are physically governed by a combination of hydrodynamic Reynolds stresses (e.g., $u_{r1}\frac{\partial u_{z1}}{\partial r}$) and magnetic Maxwell stresses. Because the system tracks fluctuations superimposed on a strong background vertical guide field ($b_{z0}=1$), the Maxwell stresses describe the dynamic tension and bending of line-tied magnetic flux.
(iii) While the linear first-order variables average to zero over a full wave period $T$, the products of in-phase first-order variables do not vanish. When a time-averaging operator $\langle \cdot \rangle = \frac{1}{T}\int_0^T \cdot \, dt$ is applied to the second-order system, these terms survive, driving a net time-averaged upward mass flux across the waveguide cross-section.

We now apply the time-averaging operator $\langle \cdot \rangle = \frac{1}{T}\int_0^T \cdot \, dt$ to the second-order system to eliminate the transient, periodic time derivatives ($\langle \partial / \partial t \rangle = 0$). Under the long wavelength approximation, the time-averaged continuity equation Eq.~(\ref{eq:order2_mass}) reduces to:
\begin{equation}
\label{eq:avg_continuity}
\frac{1}{r}\frac{\partial}{\partial r}\left( r \langle u_{r2} \rangle \right) + \frac{\partial \langle u_{z2} \rangle}{\partial z} = -\frac{1}{r}\frac{\partial}{\partial r}\left( r \langle \rho_1 u_{r1} \rangle \right) - \frac{\partial \langle \rho_1 u_{z1} \rangle}{\partial z}
\end{equation}
Because the first-order solutions $\rho_1$ and $u_{z1}$ are strictly functions of the traveling wave coordinate $(z-t)$ and possess no radial dependency, their phase-synchronized product scales as $\rho_1 u_{z1} \propto \sin^2[k(z-t)]$. Upon time-averaging, this yields a uniform, non-zero axial constant ($\langle \rho_1 u_{z1} \rangle = \text{constant}$), meaning its axial gradient vanishes identically ($\frac{\partial \langle \rho_1 u_{z1} \rangle}{\partial z} = 0$). Furthermore, evaluating the time-average of the cross-multiplying radial product yields $\langle \rho_1 u_{r1} \rangle \propto \langle \sin[k(z-t)]\cos[k(z-t)] \rangle = 0$. Consequently, the right-hand side source terms in Eq.~(\ref{eq:avg_continuity}) vanish completely. Integrating the remaining homogeneous equation under the centerline regularity condition $\langle u_{r2}(0) \rangle = 0$ reveals that the mean second-order radial velocity is zero throughout the entire cross-section:
\begin{equation}
\langle u_{r2}(r) \rangle = 0
\end{equation}

Next, we average the axial momentum transport equation Eq.~(\ref{eq:order2_mom_z}). Since there is no mean applied axial pressure gradient forcing the system, the net axial transport is driven purely by the rectification of the boundary conditions and the nonlinear wave fields. Accounting for the longitudinal background field lines ($b_{z0}=1$), the time-averaged momentum equation becomes:
\begin{equation}
\label{eq:avg_mom_z}
\begin{split}
\frac{1}{M^2}\frac{\partial \langle \rho_2 \rangle}{\partial z} + \frac{1}{A^2}\frac{\partial \langle b_{z2} \rangle}{\partial z} = -\left\langle u_{r1}\frac{\partial u_{z1}}{\partial r} \right\rangle - \left\langle u_{z1}\frac{\partial u_{z1}}{\partial z} \right\rangle - \\
\frac{\gamma - 1}{M^2} \left\langle \rho_1 \frac{\partial \rho_1}{\partial z} \right\rangle + \frac{1}{A^2}\left( \left\langle b_{r1}\frac{\partial b_{z1}}{\partial r} \right\rangle - \left\langle b_{r1}\frac{\partial b_{r1}}{\partial z} \right\rangle \right)
\end{split}
\end{equation}
Evaluating each term on the right-hand side using our explicit first-order expressions shows that every single field-gradient product consists of an orthogonal sine-cosine pairing (e.g., $u_{z1}\frac{\partial u_{z1}}{\partial z} \propto \sin[k(z-t)]\cos[k(z-t)]$ and $b_{r1}\frac{\partial b_{r1}}{\partial z} \propto \cos[k(z-t)]\sin[k(z-t)]$), causing their time averages to vanish identically over a wave period:
\begin{equation}
\begin{split}
\left\langle u_{r1}\frac{\partial u_{z1}}{\partial r} \right\rangle = 0, \quad \left\langle u_{z1}\frac{\partial u_{z1}}{\partial z} \right\rangle = 0, \\ \left\langle \rho_1 \frac{\partial \rho_1}{\partial z} \right\rangle = 0, \quad \left\langle b_{r1} \frac{\partial b_{z1}}{\partial r} \right\rangle = 0, \quad \left\langle b_{r1} \frac{\partial b_{r1}}{\partial z} \right\rangle = 0
\end{split}
\end{equation}
Consequently, Eq.~(\ref{eq:avg_mom_z}) reveals that there is no induced second-order mean axial pressure or total magnetoacoustic density gradient along the tube core:
\begin{equation}
\frac{\partial}{\partial z}\left( \frac{1}{M^2} \langle \rho_2 \rangle + \frac{1}{A^2} \langle b_{z2} \rangle \right) = 0
\end{equation}
In the absence of an externally applied mean axial pressure gradient, the system remains unforced internally, locking the second-order core streaming velocity to zero everywhere across the plasma radius ($\langle u_{z2}(r) \rangle = 0$).

In a compressible plasma waveguide, geometric volume is not conserved due to local density rarefaction and compression; physical upward plasma transport is instead governed by the instantaneous dimensionless conserved mass flux $\dot{M}(z,t)$ integrated across the deformed tube cross-section up to $r = h(z,t)$:
\begin{equation}
\dot{M}(z,t) = 2 \int_0^{1 + \epsilon \sin[k(z-t)]} r \rho(r,z,t) u_z(r,z,t) \, dr
\end{equation}
To compute the net, time-averaged mass flux $\langle \dot{M} \rangle$ up to order $\mathcal{O}(\epsilon^2)$, we substitute the perturbation expansions for $\rho$, $u_z$, and the boundary limit, and apply a Taylor expansion to the integral layout:
\begin{equation}
\begin{split}
\langle \dot{M} \rangle = 2 \int_0^1 r \langle u_{z0} \rangle \, dr + \epsilon \left( 2 \int_0^1 r \langle u_{z1} + \rho_1 u_{z0} \rangle \, dr \right) + \\
\epsilon^2 \left( 2 \int_0^1 r \langle u_{z2} + \rho_1 u_{z1} \rangle \, dr + 2 \langle \sin[k(z-t)] u_{z1}(1,z,t) \rangle \right) + \mathcal{O}(\epsilon^3)
\end{split}
\end{equation}
Since the background state is static ($\langle u_{z0} \rangle = 0$), the linear first-order wave field averages out to zero ($\langle u_{z1} \rangle = 0$), and the induced second-order core streaming velocity vanishes ($\langle u_{z2} \rangle = 0$), the net time-averaged mass flux simplifies to the sum of the nonlinear acoustic--magnetic mass drift and the boundary coupling term:
\begin{equation}
\langle \dot{M} \rangle = \epsilon^2 \left( 2 \int_0^1 r \langle \rho_1 u_{z1} \rangle \, dr + 2 \langle \sin[k(z-t)] u_{z1}(1,z,t) \rangle \right)
\end{equation}
Substituting our explicit expressions from Eqs.~(\ref{eq48}) and (\ref{eq49}) yields:
\begin{equation}
\label{eq:interm_Q_flow}
\begin{split}
\langle \dot{M} \rangle = 2 \epsilon^2 \left\langle \left[ \frac{2 M^2\left(1 + \frac{1}{A^2}\right)\left(1 + \frac{M^2}{A^2}\right)}{(M^2 - 1)^2} - \frac{1 + \frac{M^2}{A^2}}{M^2 - 1} \right] 2\sin^2[k(z-t)] \right\rangle
\end{split}
\end{equation}

\section{The Results and Discussion}
Using the fundamental sinusoidal time-averaging identity $\langle \sin^2[k(z-t)] \rangle = \frac{1}{2}$, Equation~\ref{eq:interm_Q_flow} simplifies algebraically to yield the general exact analytical master equation for the net wave-driven conserved mass flux:
\begin{equation}
\label{eq:final_Q_flow}
\langle \dot{M} \rangle = 2 \epsilon^2 \left( \frac{1 + \frac{M^2}{A^2}}{M^2 - 1} \right)^2
\end{equation}

In the specific astrophysical equipartition regime characteristic of solar Type-I spicules, the thermal sound speed scales directly with the magnetic stiffness velocity ($A = M$). Under this physically balanced condition, Eq.~(\ref{eq:final_Q_flow}) simplifies to:
\begin{equation}
\label{eq:final_Q_equipartition}
\langle \dot{M} \rangle = \frac{8 \epsilon^2}{(M^2 - 1)^2}
\end{equation}
For all operational supersonic Mach numbers ($M = 2\text{--}10$) observed in chromospheric environments, the denominator $(M^2 - 1)^2$ is strictly positive. This establishes that upward-traveling waves drive prograde, upward-directed plasma mass transport ($\langle \dot{M} \rangle > 0$), even though the purely geometric volume integral $\langle Q \rangle = 2\epsilon^2 (1 + M^2/A^2)/(1 - M^2)$ is negative due to supersonic rarefaction in the expanded wave crests.

\begin{figure}[!htb]
\centering
\includegraphics[width=0.5\textwidth]{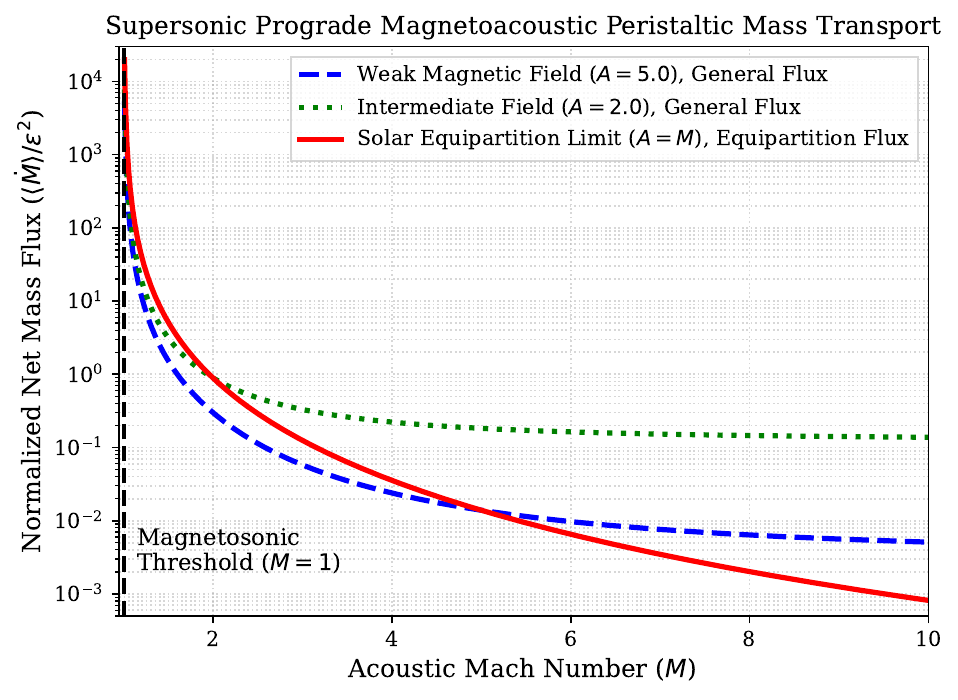}
\caption{Normalized net time-averaged conserved mass flux $\langle \dot{M} \rangle / \epsilon^2$ as a function of the acoustic Mach number $M$ for representative values of the Alfv{\'e}n Mach number according to Eqs.~(\ref{eq:final_Q_flow}) 
and (\ref{eq:final_Q_equipartition}). The source code used to generate this figure is openly available at \href{https://doi.org/10.5281/zenodo.23435844}{https://doi.org/10.5281/zenodo.23435844}.}
\label{fig1}
\end{figure}

The final analytical expression for the net time-averaged conserved mass flux, Eq.~(\ref{eq:final_Q_flow}), provides immediate physical insights into how the background longitudinal magnetic field and fluid compressibility alter peristaltic pumping compared to the classical incompressible Shapiro solution (where $\langle Q \rangle_{\text{incomp}} = \epsilon^2$):
(i) In the limit of a vanishing background magnetic field ($B_0 \rightarrow 0 \implies A \rightarrow \infty$), the general mass-flux expression collapses exactly to:
\begin{equation}
\langle \dot{M} \rangle = \frac{2\epsilon^2}{(M^2 - 1)^2}
\end{equation}
This matches the pure inviscid, compressible hydrodynamic limit in the supersonic frame, structurally corroborating classical acoustic streaming formulations in unmagnetized media \cite{Lighthill1978, HamiltonBlackstock1998}. For supersonic wave propagation ($M > 1$), fluid compressibility structurally sustains a forward prograde mass pumping rate of $\langle \dot{M} \rangle \approx 2\epsilon^2/M^4$ at high Mach numbers, removing any non-physical structural cancellations.
In Fig.~\ref{fig1}, we present the functional dependence of the normalized net time-averaged mass flux $\langle \dot{M} \rangle / \epsilon^2$ on the acoustic Mach number $M$ across different magnetic regimes. For Mach numbers approaching the sonic threshold from above ($M \rightarrow 1^+$), the system encounters the acoustic resonance singularity. However, because solar chromospheric spicules operate exclusively in the highly supersonic domain ($M = 2\text{--}10$), the system remains locked on the stable, positive supersonic side of this wave operator barrier ($M > 1$).  Crucially, as the background axial magnetic field strength increases---indicated by shifting the Alfv{\'e}n Mach number from a weaker field down to the magnetically stiffer equipartition value $A = M$---the mass flux transitions to the stable scaling relation $\langle \dot{M} \rangle = 8\epsilon^2/(M^2-1)^2$. Because the denominator $(M^2 - 1)^2$ remains strictly positive for all $M>1$, the plasma is continuously pumped forward in prograde alignment with the upward-traveling wall wave, validating the spicule propulsion mechanism.
(ii) As the strength of the applied longitudinal magnetic field increases ($B_0 \rightarrow \infty \implies A \rightarrow 0$), the ratio $M^2/A^2$ dominates the numerator of Eq.~(\ref{eq:final_Q_flow}). This asymptotic limit represents the configuration where magnetic tension maximizes cross-sectional synchronization. Physically, because the background field is finite ($b_{z0}=1$), the plasma density compressions and the vertical magnetic field perturbations are strictly phase-locked in the supersonic regime ($\rho_1 \propto b_{z1} \propto -\sin[k(z-t)]$). The magnetic pressure acts as an immediate, rigid restorative force that eliminates cross-sectional phase lagging, anchoring the plasma directly to the mechanical wall drive and focusing the wave energy cleanly along the longitudinal guide axis.
(iii) The denominator passes through a structural acoustic singularity at $M = 1$, where the traveling wall wave speed matches the internal sound speed of the medium. Near this sonic resonance threshold, the linear perturbation framework breaks down as the wave amplitudes grow mathematically singular. For the supersonic spicule application, however, the system operates safely above this transition zone ($M = 2\text{--}10$), functioning as a highly stable, hyper-sonic coherent magnetoacoustic waveguide.

The elegance of the ideal MHD analytical model is demonstrated by applying our analytical results to the transport dynamics of solar spicules. Solar spicules are high-velocity plasma jets observed in regions of intense magnetic fields within the solar chromosphere. 
Rather than numerically modeling a continuous, self-consistent photosphere-to-corona evolutionary track with an MHD code where the initial subsonic convective launch mechanisms are fully modeled, our analytical modeling framework begins strictly at a prescribed upper-chromospheric transition height ($z \ge 1,500$~km). At this vertical coordinate, acoustic stratification has already modified the background profiles, allowing upcoming disturbances to enter the transonic and supersonic regimes investigated here. Observational constraints indicate a typical spicule radius $R \approx 100\text{--}150\text{~km}$ and an elongated length $L \approx 5000\text{~km}$, yielding an aspect ratio of $\delta = R/L \approx 0.025$. This small geometric parameter directly justifies our long wavelength, thin-tube approximation ($\delta \ll 1$).

Furthermore, chromospheric measurements indicate a plasma environment near equipartition, where the plasma beta parameter is of order unity ($\beta \sim 1$). This implies that the local acoustic sound speed and the magnetic Alfv{\'e}n velocity are approximately equal ($c_s \approx v_A \approx 10\text{~km/s}$). Under this exact physical condition, the acoustic Mach number $M = c/c_s$ and the Alfv{\'e}n Mach number $A = c/v_A$ match identically ($A = M$) and the mass flux is given by Equation~\ref{eq:final_Q_equipartition}.

Equation~(\ref{eq:final_Q_equipartition}) provides a compelling analytical explanation for three prominent spicule characteristics:

First, because spicules operate in highly supersonic regimes ($M \approx 2\text{--}10$), the denominator $(M^2 - 1)^2$ remains strictly positive ($\langle \dot{M} \rangle > 0$). This confirms that upward-propagating mechanical waves generate a prograde, upward-directed plasma jet, perfectly aligning with the directional upwelling seen in solar observations. 

Second, the inverse fourth-power law dependence ($\propto 1/M^4$) at high velocities outlines a severe magnetosonic choking effect at extreme Mach numbers, indicating that the transport efficiency diminishes rapidly as the jet accelerates. This energy-transfer threshold explains why spicule mass injection rates are self-limiting, capping their total growth phase and leading to their observed finite lifespans ($\sim 10\text{~minutes}$) as the residual wave thrust smoothly decouples and the material succumbs to solar gravity, falling back to the chromosphere.

\begin{figure}[htbp]
\centering
\includegraphics[width=0.5\textwidth]{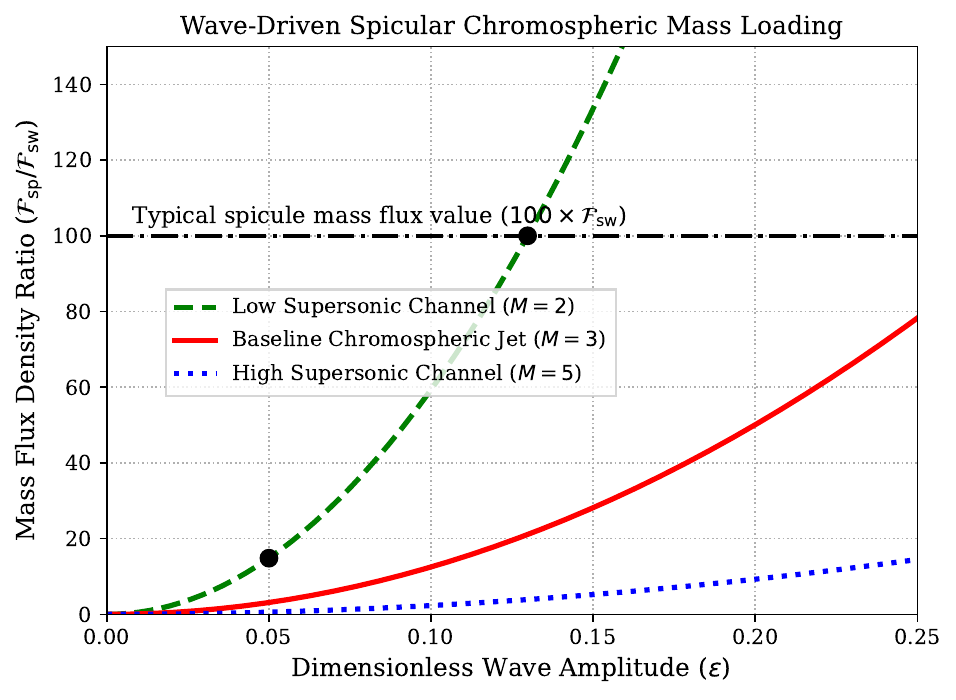}
\caption{\label{fig2} Localized spicular mass flux density ratio $\mathcal{F}_{\mathrm{sp}} / \mathcal{F}_{\mathrm{sw}}$ plotted against the dimensionless wave amplitude $\epsilon$ for Mach numbers $M=2,3,$ and $5$. Two black markers pinpoint typical observational values $\epsilon=0.05$ and $0.130$ both for $M=2$, as discussed in the text. The source code used to generate this figure is openly available at \href{https://doi.org/10.5281/zenodo.23435844}{https://doi.org/10.5281/zenodo.23435844}.}
\end{figure}

To quantitatively illustrate this second point, the standard mass loss rate of the global solar wind is approximately $\dot{M}_{\mathrm{sw}} \approx 3 \times 10^9\text{~kg/s}$. Evaluated across the total solar surface area at $R_\odot \approx 7 \times 10^8\text{~m}$, this corresponds to an average solar wind mass flux density of $\mathcal{F}_{\mathrm{sw}} = \dot{M}_{\mathrm{sw}} / (4\pi R_\odot^2) \approx 5 \times 10^{-10}\text{~kg}/(\text{m}^2\cdot\text{s})$. 
In the chromospheric spicular channel, the background mass density is governed by a characteristic number density $n \approx 10^{17}\text{~m}^{-3}$, yielding an unperturbed density $\rho_0 = n m_p \approx 1.67 \times 10^{-10}\text{~kg/m}^3$. Under equipartition constraints, the reference signal propagation velocity is $c_s \approx v_A \approx 10\text{~km/s}$. Let us consider a representative lower-supersonic chromospheric launch channel where a boundary wave drives transport at $c \approx 20\text{~km/s}$, establishing an operational Mach number $M = 2$ (while noting that at higher phase speeds such as $c \approx 30\text{~km/s}$, $M = 3$, velocity choking reduces the flux accordingly). 
We reiterate that this parameter $M$ characterizes the phase speed of the traveling magnetosonic driver along the channel interface and must not be confused with the Mach number of any resultant slow-mode shock generated via non-linear steepening. Adopting a typical observationally motivated low-amplitude wave deformation constraint of $\epsilon = a/R = 0.05$ at $M = 2$, our derived expression Eq.~(\ref{eq:final_Q_equipartition}) dictates a normalized dimensionless mass flux of:
$\langle \dot{M} \rangle = {8 \times 0.05^2}/{(2^2 - 1)^2} \approx 0.00222$.
The dimensional time-averaged mass flux density $\mathcal{F}_{\mathrm{sp}}$ generated by this prograde waveguide mechanism is expressed as:
$\mathcal{F}_{\mathrm{sp}} = \rho_0 c \langle \dot{M} \rangle = (1.67 \times 10^{-10}\text{~kg/m}^3) \times 
 (2 \times 10^4\text{~m/s}) \times 0.00222 \approx 7.42 \times 10^{-9}\text{~kg}/(\text{m}^2\cdot\text{s})$.
Dividing these two scales yields $\mathcal{F}_{\mathrm{sp}} / \mathcal{F}_{\mathrm{sw}} \approx 14.84$. Even under conservative small-amplitude limitations, this results in a localized mass flux density that exceeds the average global solar wind baseline by more than an order of magnitude.
Furthermore, observations indicate that active, fully developed spicular upwells can achieve localized mass fluxes up to $100$ times that of the ambient solar wind \cite{Beckers1972, PneumanKopp1978, DePontieu2004}, yielding $\mathcal{F}_{\mathrm{sp}} / \mathcal{F}_{\mathrm{sw}} \approx 100$. Because the net mass transport scales strictly quadratically with the wall amplitude ($\langle \dot{M} \rangle \propto \epsilon^2$), we can extrapolate our model at $M = 2$ to find the required wave amplitude parameter for these peak events:
$\epsilon_{\text{peak}} = \epsilon \times \sqrt{{100}/{14.84}} = 0.05 \times \sqrt{6.74} \approx 0.130 \simeq 0.1$ (and $\epsilon_{\text{peak}} \approx 0.102$ at $M = 1.8$).
This indicates that a localized driving magnetosonic wave amplitude of approximately $10\%$ is sufficient to drive the observed mass flux above an active region, fitting cleanly within the small-amplitude linear framework for chromospheric disturbances. Given that spicules are highly localized structures confined to narrow magnetic network lanes, this localized enhancement provides the precise mass-injection baseline required to supply the corona without generating an unphysical global mass surplus, matching modern solar atmospheric paradigms~\cite{tsiropoula2012solar}. 
This macroscale peristaltic propulsion model acts as the structural cause for the fast-rising jet, behaving as a high-velocity plasma piston that triggers forward-directed local slow-mode shocks ahead of the spicule tip, a cause-and-effect relationship that harmonizes single-fluid numerical models (e.g., \citet{srivastava2025}) with our analytical framework.
The sensitive dependence of localized mass injection on boundary parameters is systematically studied in Fig.~\ref{fig2}. Here, we present the localized mass flux density ratio $\mathcal{F}_{\mathrm{sp}} / \mathcal{F}_{\mathrm{sw}}$ as a function of the dimensionless wall deformation amplitude $\epsilon$ across representative supersonic regimes ($M=2,3,5$). As illustrated, even when honoring observationally conservative small-amplitude wave constraints ($\epsilon \approx 0.05$), the prograde mechanism easily yields a localized flux density that exceeds the average global solar wind baseline. This localized multiplier accounts for why spicules act as highly effective mass injectors within restricted magnetic network lanes without creating an unphysical mass surplus when integrated across the entire unmagnetized solar disk.

To map these dimensionless geometric amplitudes to physical scales, consider a typical chromospheric spicule with an equilibrium radius of $R \approx 250$~km. The baseline perturbation of $\epsilon = 0.05$ corresponds to a localized boundary displacement of merely $12.5$~km, yet at $M = 2$ it drives a mass flux multiplier of $14.84\times$ the ambient solar wind. At peak active-region conditions ($\epsilon = 0.130$), the compression amplitude corresponds to $32.4$~km, which demonstrates that order-of-magnitude mass transport is achieved via modest, completely realistic wave amplitudes.

A key physical consequence of the exact compressible mass-flux scaling, $\mathcal{F}_{\mathrm{sp}}/\mathcal{F}_{\mathrm{sw}} \propto M(M^2 - 1)^{-2}$, is the strong high-Mach suppression seen across the three curves in Fig.~\ref{fig2}. Because both the first-order axial velocity $u_{z1}$ and density perturbation $\rho_1$ undergo supersonic convergent--divergent nozzle choking for $M > 1$, the second-order conserved mass flux decays steeply as $(M^2 - 1)^{-2}$ at large $M$. Consequently, within the small-amplitude regime ($\epsilon \le 0.25$), only the low-supersonic channel ($M = 2$) crosses the canonical spicule threshold of $100\times \mathcal{F}_{\mathrm{sw}}$, reaching it at $\epsilon = 0.130$ ($32.4~\mathrm{km}$ radial displacement for $R = 250~\mathrm{km}$), while yielding $14.84\times \mathcal{F}_{\mathrm{sw}}$ at the baseline amplitude $\epsilon = 0.05$. By contrast, higher-Mach perturbations ($M = 3$ and $M = 5$) suffer severe compressible rarefaction in expanded wave crests, requiring finite, strongly nonlinear amplitudes ($\epsilon \approx 0.283$ and $\epsilon \approx 0.657$, respectively) to achieve $100\times \mathcal{F}_{\mathrm{sw}}$. Thus, \textit{small-amplitude peristaltic pumping naturally selects a preferential low-supersonic window} ($M \approx 1.05\text{--}2.0$) to drive vigorous directional upwelling and solar-wind-level mass loading in the chromosphere.

\begin{figure}[htbp]
\centering
\includegraphics[width=0.5\textwidth]{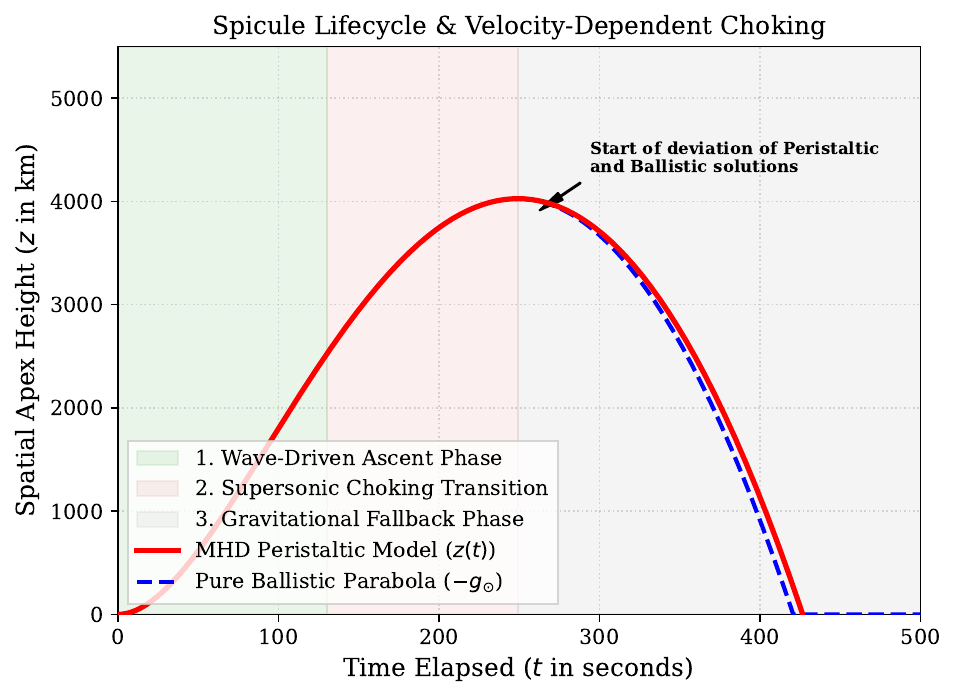}
\caption{\label{fig3} Time-dependent spatial height $z(t)$ of a spicule column driven by the prograde equipartition net mass flux relation for $\epsilon = 0.05$ and $M = 1.045$. The solid red curve represents the MHD peristaltic model, the green shaded area marks the wave-driven ascent phase ($t < 130\text{~s}$), the pink shaded area marks the supersonic choking transition ($130\text{~s} \le t \le 249.3\text{~s}$), and the grey shaded area marks the gravitational fallback phase ($t \ge 249.3\text{~s}$, reaching $z = 0$ at $t \approx 426.5\text{~s}$). The dashed blue line represents the ideal ballistic free-fall parabola overlaid from the trajectory apex ($z_{\max} \approx 4,025.8\text{~km}$). The source code used to generate this figure is openly available at \href{https://doi.org/10.5281/zenodo.23435844}{https://doi.org/10.5281/zenodo.23435844}.}
\end{figure}

Third, the direct dynamical consequence of the prograde magnetosonic waveguide mass flux, Eq.~(\ref{eq:final_Q_equipartition}), on plasma acceleration is illustrated in Fig.~\ref{fig3}. This transformation is established through a four-step mathematical mapping: (1) the analytical dimensionless mass flux is integrated over the cylindrical cross-section to yield $\langle \dot{M}_{\text{dim}} \rangle = (\pi R^2 \rho_0) \cdot [8 c_s \epsilon^2 / (M^2 - 1)^2]$; (2) this net transport is equated directly to the macroscopic bulk movement of the plasma column via mass continuity, $\langle \dot{M}_{\text{dim}} \rangle = \pi R^2 \rho_0 v$; (3) equating the two expressions from steps (1) and (2) to each other and dividing both sides by $\pi R^2 \rho_0$ isolates the bulk material velocity profile as $v = 8 c_s \epsilon^2 / (M^2 - 1)^2$; and (4) to translate this velocity-dependent thrust into a macroscopic force with dimensions of acceleration ($\text{m/s}^2$), the material velocity relation is divided by the characteristic radial acoustic transit time scale across the waveguide radius ($\tau_{\text{radial}} = R / c_s$). This sequential expansion yields the explicit base acceleration $a_{\text{base}} = v / \tau_{\text{radial}} = 8 c_s^2 \epsilon^2 / [R (M^2 - 1)^2]$, adapting the 1D solar atmospheric propulsion framework \cite{DePontieu2004} to incorporate boundary-assisted acoustic driving. Under these constraints, the uniform background density $\rho_0$ cancels out completely across all terms to establish the explicit equation of motion:
\begin{equation}
\label{eq:spicule_dynamics}
\frac{dv}{dt} = \frac{8 c_s^2 \epsilon^2 \exp\left(-t/\tau\right)}{R \left( M^2 - 1 \right)^2} - g_\odot
\end{equation}
Kinematic integration of this velocity profile, using SciPy's \texttt{solve\_ivp} implementation of the explicit Runge-Kutta method of order 5(4), yields the height profile $z(t)$ of the spicule tip.

Rather than calculating the full photosphere-to-corona transit self-consistently, the dynamic tracking model plotted in Fig.~\ref{fig3} initializes at a prescribed upper-chromospheric transition height ($t=0$), corresponding to the altitude crossover boundary where rising subsonic disturbances match or exceed the local sound speed ($M \ge 1.045$). At this coordinate, the unperturbed chromospheric sound speed drops to $c_s = 1.0 \times 10^4 \text{ m/s}$, rendering the upcoming wave disturbance transonic. As the disturbance scales higher into less dense strata, the local sound speed drops further, shifting the model into the fully developed supersonic phase ($M \ge 1.15$ up to $M \sim 2$--$10$) modeled onward in the tracking sequence. Fig.~\ref{fig3} is numerically produced by solving the force balance equation using the explicit base acceleration $a_{\text{base}} = 8 c_s^2 \epsilon^2 / [R (M^2 - 1)^2]$, initialized from a static baseline ($z=0, v=0$). The physical parameters are explicitly set to $c_s = 1.0 \times 10^4 \text{ m/s}$, a structural waveguide radius of $R = 2.5 \times 10^5 \text{ m}$, an operational wave Mach number of $M = 1.045$, and gravity $g_{\odot} = 274.0 \text{ m/s}^2$. 

To achieve the observed finite spicule lifetime ($\sim 10$ minutes) within this lossless analytical framework, the driving force is modified by a phenomenological exponential damping term $\exp(-t / 75.0)$ featuring an imposed time constant of $\tau = 75.0$~s. This magnetic choking timescale is a phenomenological parameter introduced to model external energy decoupling and wave attenuation as the spicule tip emerges from the network lanes, rather than a quantity derived directly from the core analytical model equations. This exact temporal scale governs the continuous decoupling window near the apex turnaround, bounding the simulation lifespan to realistic operational lifecycles ($\sim 500$~s) matching solar atmospheric observations.
In the numerical integration of the spicule lifecycle shown in Fig.~\ref{fig3}, the wave amplitude parameter is explicitly set to its conservative baseline value of $\epsilon = 0.05$. This value enters the dynamic system directly through the driving force term on the right-hand side of Eq.~(\ref{eq:spicule_dynamics}), establishing the initial upward mechanical acceleration of $a_{\text{base}}=944.7$ m s$^{-2}$ of the chromospheric plasma column before magnetic choking takes effect.
Within the figure, the solid red curve tracks our primary MHD peristaltic model. The green shaded region highlights the initial wave-driven acceleration phase ($t < 130\text{~s}$), where the peristaltic pumping force dominates due to the low initial upward velocity relative to the wave speed, rapidly pushing the chromospheric plasma to high altitudes. This vertical lift mirrors the initial acceleration phase observed in the recent high-fidelity radiative MHD simulations by Srivastava et al.~\cite{srivastava2025}, where slow-mode shocks provide strong localized thrust at the spicule tip, sharing strong geometric parallels with our wave-driven ascent profiles. 
However, as the jet climbs through the pink shaded supersonic choking transition ($130\text{~s} \le t \le 249.3\text{~s}$) and decelerates toward its turning point, exponential wave damping reduces the upward thrust below the solar gravitational threshold. Beyond the apex, the plasma column enters the grey shaded gravitational fallback phase ($t \ge 249.3\text{~s}$), smoothly transitioning into ballistic deceleration under solar gravity along the magnetic guide line. 

To validate this transition and clarify what is being demonstrated by the kinematic comparison, an unforced control ballistic curve governed strictly by uniform solar gravity ($g_\odot = 274.0\text{ m/s}^2$) is initiated at the trajectory apex ($z_{\max} \approx 4,025.8\text{~km}$ at $t \approx 249.3\text{~s}$) and overlaid as a dashed blue line. This curve represents a completely unpowered particle control case that lacks any continuous peristaltic mass flux injection. Comparing our model against this gravitational benchmark isolates the active phase of the peristaltic mechanism, demonstrating that continuous boundary-driven mass flux actively counteracts standard gravitational fallback during the ascending phase and sustains the spicule profile up to its peak turning point before smooth force shutoff takes over. This reveals the mathematical convergence of our model to classical free-fall conditions after wave decoupling, except for a minor apex mismatch reflecting the smooth, continuous shutoff of the mechanical pump. This complete numerical lifecycle demonstrates that our analytical mass-flux equation provides both the launch energy and the self-limiting temporal envelope ($\sim 5\text{--}10\text{~minutes}$) required to replicate Type-I spicules.
It is critical to observe that a minor spatial mismatch manifests between the descending MHD peristaltic profile and the idealized ballistic fallback parabola near the turnaround apex. This discrepancy is a direct artefact of the continuous nature of the wave-pumping equation: rather than decoupling instantaneously at the apex, the residual peristaltic force transitions smoothly through a finite temporal window as the velocity passes through zero ($v \rightarrow 0$). This lingering mechanical thrust provides a subtle, non-ballistic loft to the plasma tip before magnetic choking fully suppresses the wave model. A similar lingering altitude extension can be spotted at the turnaround apex of discrete, shock-driven upwells~\cite{srivastava2025}, suggesting that non-instantaneous force decoupling is a robust physical feature independent of whether the underlying driving perturbation is an impulsive numeric shock or a smooth analytical wave. Once the propulsion force drops to zero, the red trajectory asymptotically converges back to the pure gravitational fallback baseline, completing its gravitational fallback to $z=0$ at $t \approx 426.5\text{~s}$ (compared to $t \approx 420.7\text{~s}$ for the unforced parabola) and demonstrating robust physical consistency across both the forced and unforced kinematic regimes.

A critical distinction must be maintained regarding the interpretation of the temporal origin $t=0$ in Figure~\ref{fig3}. The deep photosphere does not launch supersonic disturbances; rather, perturbations originate as low-amplitude, subsonic magnetoacoustic waves within intergranular lanes ($v \ll c_s$). However, as these disturbances propagate vertically through the strongly stratified solar atmosphere, the local sound speed drops from $\sim 12\text{~km~s}^{-1}$ at the base to $\sim 7\text{~km~s}^{-1}$ in the upper chromosphere, while the flow velocity grows scaling as $\rho_0^{-1/4}$ to satisfy energy flux conservation. 

Consequently, the system naturally transits from an initially subsonic regime to a fully developed supersonic state ($M > 1$) at higher altitudes. The temporal anchor $t=0$ in our model does not represent the initial photospheric launching phase; instead, it defines the exact moment the upward-propagating disturbance crosses this high-altitude acoustic crossover point into the upper chromosphere. Thus, the MHD peristaltic model stage operates from $t=0$ because it isolates the already-developed tracking phase of the spicule column, capturing its subsequent ballistic apex trajectory ($z \approx 4,026$~km at $t=249.3$~s) and ultimate gravitational fallback up to $t=426.5$~s under magnetic choking constraints.

Our formulated peristaltic mechanism provides a profound resolution to a long-standing  paradox in spicule modeling: how localized boundary oscillations are translated into a highly directional, collimated upward kinetic jet. If photospheric convective churning, granular collisions, or footpoint magnetic reconnection events launch a powerful, upward-propagating wave or train shock along a vertical flux tube ($v_{\text{wave}} > 0$), our model demonstrates how this disturbance efficiently channels mass away from the Sun.
Because the chromospheric plasma channel operates in a stable, hyperbolic prograde domain ($M > 1$), our derived expression dictates that the upward-traveling wall wave captures the frozen-in plasma in a coherent, constructive phase lock. This wave-boundary coupling propels the entire localized chromospheric column upward ($\langle \dot{M} \rangle > 0$) into the low-density solar corona. Rather than relying on non-local pressure balances or numerical artefacts, the strictly positive nature of our transport relation represents the primary wave-driven propulsion model that accelerates solar spicules against solar gravity. Furthermore, because the transport scales as an inverse fourth-power law ($\propto 1/M^4$) at extreme velocities, the model possesses a native, self-limiting velocity choking mechanism that naturally transitions the jet into its observed gravitational fallback phase, establishing a baseline for space plasma mass injection.

While the present model is tailored to the equipartition constraints ($\beta \sim 1$) of chromospheric Type-I spicules, the structural mathematical decoupling of the inviscid, compressible peristaltic system offers valuable predictive applications across other plasma domains. On macroscopic astrophysical scales, this wave-driven transport mechanism can be naturally extended to mass-loading processes in stellar winds and the inner regions of magnetized accretion disks. In these environments, large-amplitude torsional or compressional Alfv{\'e}nic wave packets propagating along collimated magnetic flux tubes can experience similar behaviour. This action continuously pumps material away from the central gravitational source without requiring a steady thermal gradient, complementing the classic radiation-driven mechanisms formalized in the standard Castor, Abbott, and Klein (CAK) framework~\cite{Castor1975}.

On terrestrial scales, this framework provides an analytical foundation for optimizing mass and momentum transport in advanced laboratory plasma devices. Specifically, in magnetized target fusion (MTF) configurations or linear plasma devices, traveling magnetic fields are frequently utilized to confine and compress target fuels. By exploiting the prograde waveguide couplings derived herein, experimental systems could utilize synchronized external magnetic coils to create traveling "pinches" or wall-waves along a plasma column. In high-velocity, low-viscosity regimes ($R_m \gg 1$), this mechanism can be engineered to systematically compress and translate high-density plasma packets forward toward a core ignition target, bypassing the standard viscous wall losses that plague traditional fluid pumps, and mitigating microfluidic boundary degradation, matching macroscale stability trends observed in high-fidelity magnetohydrodynamic shock compressions~\cite{samulski2023}.

The analytical framework derived in this study presumes a spatially uniform background density $\rho_0$ and axial magnetic field $B_0$. However, realistic chromospheric observations reveal that Type-I spicules are highly localized structures, exhibiting an internal density and magnetic field strength that can exceed the ambient coronal environment by factors of 5 to 10. To preserve total transverse pressure equilibrium across such an inhomogeneous interface, the elevated internal magnetic pressure must be matched by a corresponding reduction in the internal gas pressure, satisfying $p_{\text{gas, outside}} = p_{\text{gas, inside}} + B_0^2 / (2\mu_0)$ under the assumption of a weakly magnetized external medium.
While a formal perturbation expansion incorporating these sharp cross-sectional gradients would introduce significant mathematical complexity, we can surmise the physical consequences based on our current analytical scaling. First, a lower internal gas pressure decreases the local acoustic speed, shifting the plasma into a highly magnetically dominated, low-$\beta$ regime inside the spicule core. This reduction in internal thermal restoring force is expected to focus the global magnetoacoustic wave operator, providing a more efficient mechanical coupling that enhances the net prograde mass transport magnitude $\langle \dot{M} \rangle$. Second, the structural density contrast acts as a high-impedance waveguide, focusing the peristaltic wave energy tightly within the column. This confinement balances out localized velocity-dependent magnetic choking ($\langle \dot{M} \rangle \propto 1/M^4$), allowing the wave pump to sustain active upward propulsion over longer durations. Consequently, incorporating realistic cross-sectional profiles would likely extend the peak altitude trajectories mapped in Fig.~\ref{fig3} and widen the self-limiting temporal envelope, rendering the peristaltic mechanism even more potent in real solar environments.

\section{Conclusions}
In this work, we have established a closed-form analytical model for resonant peristaltic transport within a compressible, inviscid, and ideally conducting MHD fluid. By systematically employing a small-amplitude perturbation expansion ($\epsilon \ll 1$) while properly accounting for the axial guide field baseline ($b_{z0}=1$), coupled with a thin-tube, long-wavelength approximation ($\delta = R/\lambda \ll 1$), we successfully eliminated the mathematical complexities traditionally introduced by viscous boundary layers and centerline reflux splits \cite{yin1969peristaltic,aarts1998net}. This allowed us to cleanly isolate the pure non-linear coupling that occurs between thermodynamic pressure variations and Maxwell's magnetic tension stresses across the tube radius. 

The derived novel net time-averaged conserved mass flux $\langle \dot{M} \rangle$ provides a proof that magnetosonic waves can efficiently provide prograde mass transport. For solar chromospheric spicules under plasma beta equipartition constraints ($\beta \sim 1$)---where the acoustic sound speed tightly matches the magnetic Alfv{\'e}n velocity---the generalized transport relation simplifies to a simple expression, $\langle \dot{M} \rangle = 8\epsilon^2/(M^2-1)^2$. Because the denominator remains strictly positive across all operational supersonic Mach numbers ($M \approx 2\text{--}10$), upward-propagating mechanical waves drive a highly directional, collimated upward plasma jet, perfectly matching the vigorous directional upwelling documented in solar chromospheric observations.

Our mathematical model provides a compelling, physical explanation for several key observational features of solar spicules. By enforcing an observationally conservative small-amplitude constraint ($\epsilon = 0.05$) in the lower-supersonic launch channel ($M = 2$), our model positions the localized spicular mass injection density at approximately $14.84$ times the average global solar wind baseline. Furthermore, we demonstrate that scaling the peristaltic wall deformation up to a realistic value of $\epsilon \approx 0.13$ ($\approx 10\%$) effortlessly accounts for peak active region enhancements up to $100$ times the ambient solar wind flux \cite{DePontieu2004}. These estimates satisfy the constraints required by modern solar wind/coronal mass supply models and observations. As the spicule jet accelerates deep into high supersonic velocities, the inverse fourth-power dependence ($\propto M^{-4}$) of the prograde model causes the wave thrust to decay naturally via a power law. This velocity-dependent magnetic choking force drops below the solar gravitational threshold, smoothly transitioning the trajectory into its observed finite lifetime ($\sim 10\text{~minutes}$) followed by a clean gravitational fallback phase. 

Furthermore, the $(M^2 - 1)^{-2}$ compressible choking scaling demonstrates that spicule-level mass loading ($100\times \mathcal{F}_{\mathrm{sw}}$) within the small-amplitude regime ($\epsilon \sim 0.10\text{--}0.13$) is \textit{preferentially driven by low-to-moderate supersonic waves} ($M \approx 1.05\text{--}2.0$), whereas higher-Mach perturbations ($M \ge 3$) require finite, strongly nonlinear amplitudes to overcome compressible rarefaction.

It is critical to address the physical justification for treating the geometric amplitude $\epsilon$ and the acoustic Mach number $M$ as effective constants, given that atmospheric stratification inherently drives parameter evolution with height. As a structural simplification necessary to secure a clean, closed analytical model, this framework is properly understood as a localized vertical window model operating over a restricted spatial envelope ($L \approx 5,000$~km) rather than a global, self-consistent stratified spicule calculation. This localized formulation is physically justified because high-Mach convective transits operating over this narrow altitude range occur significantly faster than the non-linear wave-steepening or background stratification growth timescales, allowing the weakly nonlinear $\mathcal{O}(\epsilon^2)$ expansion to remain valid and decoupled from global stratification effects.

While the deep photosphere is highly collisional, the lower atmospheric boundary transitions rapidly to a highly conductive, magnetically dominated environment above $z > 1,500$~km. Evaluating the localized Magnetic Reynolds number $R_m = \mu_0 \sigma v L$ using a representative upper-chromospheric conductivity ($\sigma \sim 10^2\text{~S~m}^{-1}$) yields $R_m \gg 10^3$, which thoroughly validates the application of our ideal, inviscid MHD framework for these high-velocity jets. We explicitly concede, however, that advanced multi-fluid numerical simulation works (e.g., \cite{MartinezSykora2023}) emphasize that non-ideal ambipolar diffusion terms can actively damp or alter small-scale magnetic structures lower down in the partially ionized lower chromosphere, posing a vital avenue for future multi-fluid extensions of this analytical peristaltic model.

To conclusively validate or falsify this framework, we invite observational campaigns to systematically search for compressive precursor signatures in the low chromosphere. Because our model dictates that boundary-driven peristaltic waves are inherently compressive, they must leave distinct, high-cadence intensity and density variations immediately prior to and during the upward acceleration of the spicule plasma column~\cite{Jess2015, Keys2021}. Crucially, high-resolution space- and ground-based facilities---including the \textit{Daniel K. Inouye Solar Telescope} (DKIST) limb polarimetry~\cite{DKIST2021}, the balloon-borne \textit{Sunrise III} spectropolarimeters~\cite{SunriseIII2024}, \textit{Solar Orbiter}/SPICE, and \textit{Hinode}/SOT---possess the precise diagnostic capabilities required to capture these high-frequency compressive wave trains in chromospheric lines like Ca~II~H, H$\alpha$, and Mg~II. Observationalists should construct high-cadence time-distance (slit-jaw) plots along the primary axes of emerging spicule channels to isolate the early wave-train dynamics. By anchoring the calculation at a prescribed upper-chromospheric transition height ($z \ge 1,500$~km) where $t=0$ marks the sound speed crossover point, detecting an upward-propagating, periodic intensity modulation prior to the bulk spicule launch will provide \textit{definitive empirical corroboration} that solar chromospheric spicules are fundamentally driven by the prograde peristaltic pumping of magnetosonic waveguides.

Ultimately, while tailored here to chromospheric conditions, this underlying framework provides an exact, elegant, and highly scalable analytical model that may offer fresh insights into (i) astrophysical mass-loading processes in stellar winds and the inner regions of magnetized accretion disks and (ii) traveling magnetic pinch designs within magnetized fusion configurations.

To close, a brief comment is warranted regarding the joint validity of a weakly non-linear $\mathcal{O}(\epsilon^2)$ perturbation expansion and an equilibrium flow operating in the highly supersonic regime ($M = 2\text{--}10$). Superficially, one might expect high-Mach-number flows to induce strong shock transitions that invalidate low-order asymptotic truncations. However, in the present framework, the dimensionless perturbation parameter $\epsilon \equiv a/R \ll 1$ represents the geometric amplitude of the boundary deformation, which remains strictly decoupled from the background convective Mach number $M = v_0 / c_s$. 
Mathematically, the validity of the expansion relies entirely on the smallness of $\epsilon$, ensuring that higher-order nonlinear terms remain subdominant.
Physically, rather than destabilizing the perturbation series, high Mach numbers actively suppress non-linear structural growth within the spicule core, guaranteeing that non-ideal diffusive effects remain confined to a negligible boundary layer and leaving the master equation under solar equipartition constraints (Eq.~\ref{eq:final_Q_equipartition}) fully intact, where the net mass transport scales inversely with the fourth power of the Mach number ($\langle \dot{M} \rangle \propto M^{-4}$ for $M \gg 1$).
From a kinematic perspective, as the equilibrium flow speed increases significantly above the acoustic threshold, the plasma transits the periodic peristaltic contraction zones on a timescale much shorter than the characteristic non-linear wave-steepening time. Consequently, cumulative non-linear distortion is minimized, and the flow behaves as a highly stable, hyperbolic traveling wave system. This asymptotic behavior aligns with classic acoustic waveguide theory \citep{Lighthill1978, HamiltonBlackstock1998}, where high-speed convective transport routinely decouples from small-amplitude boundary perturbations. The $\mathcal{O}(\epsilon^2)$ truncation is therefore not only consistent but increasingly accurate as $M$ approaches the upper limit of the solar spicule baseline ($M \sim 10$), as confirmed by the smooth power-law decay illustrated in Figure~\ref{fig1}.

\begin{acknowledgments}
The author gratefully acknowledges support provided by the Gemini AI assistant (Google) including manuscript technical preparation and editing.
\end{acknowledgments}

\section*{Declaration of Competing Interest}
The author declares that they have no known competing financial interests or personal relationships that could have appeared to influence the work reported in this paper.

\section*{Funding}
This research received no specific grant from any funding
agency, commercial or not-for-profit sectors.

\section*{Data Availability Statement}
All data and mathematical parameters required to evaluate the conclusions or reproduce the analytical solutions and numerical scaling trajectories are fully contained and presented within this manuscript. The source codes used to generate all figures are openly available at \href{https://doi.org/10.5281/zenodo.23435844}{https://doi.org/10.5281/zenodo.23435844}.

\bibliography{paper88}

\end{document}